\documentclass[twocolumn,showpacs,aps,prd,nobibnotes,
  nofootinbib,floatfix,superscriptaddress,longbibliography]{revtex4-1}

\usepackage{natbib}
\usepackage{amsmath,amssymb,bm,xfrac}
\usepackage{graphicx}
\usepackage[usenames]{xcolor} 
\usepackage[normalem]{ulem}
\usepackage{xspace}
\usepackage{dsfont}
\usepackage{soul} % Strike-throughs and underline support
\usepackage{dsfont}
\usepackage{multirow}  
\usepackage{float}  
\usepackage{cancel}
\usepackage[caption=false]{subfig}
\usepackage[colorlinks=true,allcolors=blue]{hyperref}

\usepackage[caption=false]{subfig}

\newcommand{\orcid}[1]{\href{https://orcid.org/#1}{\includegraphics[height=\fontcharht\font`\B]{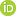}}}
\newcommand{\aap}{Astron. Astrophys.}
\newcommand{\aj}{Astron. J.}
\newcommand{\apjl}{Astrophys. J. Lett.}

\newcommand{\jcap}{J.~Cosmology Astropart. Phys.}

\newcommand{\mnras}{Mon. Notices Royal Astron. Soc.}

\newcommand{\pasp}{Publ. Astron. Soc. Pac.}

\newcommand{\beq}{\begin{equation}}
\newcommand{\eeq}{\end{equation}}

\renewcommand{\eqref}[1]{Eq.\,(\ref{#1})\xspace}

\definecolor{darkred}{cmyk}{0,1,1,0.4}

\begin{document} 

%% Paper title
    \title{
    %A novel approach to relieving the Hubble Tension using Cosmological observation. \\ OR \\
    Exploring the Hubble Tension: A Novel Approach through Cosmological Observations}

%% Author list
% First Author
\author{Darshan Kumar~\orcid{0000-0001-6665-8284}}
\email{darshanbeniwal11@gmail.com}
\affiliation{Department of Physics and Astrophysics, University of Delhi, Delhi 110007, India}

% Second Author
\author{Debajyoti Choudhury~\orcid{0000-0002-8124-0043}}
\email{debchou.physics@gmail.com}
\affiliation{Department of Physics and Astrophysics, University of Delhi, Delhi 110007, India}

% Third Author
\author{Debottam Nandi~\orcid{0000-0002-7719-9623}}
\email{dnandi@physics.du.ac.in}
\affiliation{Department of Physics and Astrophysics, University of Delhi, Delhi 110007, India}

\begin{abstract}
  The simplest cosmological model ($\Lambda$CDM) is well-known
  to suffer from the Hubble tension, namely an almost $5 \sigma$
  discrepancy between the (model-based) early-time determination of
  the Hubble constant $H_0$ and its late-time (and model-independent)
  determination.  To circumvent this, we introduce an additional
  energy source that varies with the redshift as $(1 + z)^n$,
  where $0 < n < 3$, and test it against the Pantheon Compilation of
  Type Ia Supernovae as well as the CMBR observations (at $z \approx 1100$).
  The deduced $H_0$ is now well-consistent with the value obtained
  from local observations of Cepheid variables. Suggesting
  a non-zero value for the curvature density parameter,
  positive (negative) for $n > 2$ ($n < 2$), the resolution is also
  consistent with the BAO data.

\end{abstract}

\maketitle 

%%%%%%%%%%%%%%%%%%%%%%%%%%%%%%%
%%% Section 1: Introduction %%%
%%%%%%%%%%%%%%%%%%%%%%%%%%%%%%%

\section{Introduction} 
The Hubble constant $H_0$ has been used for almost $90$ years to
describe the current rate of expansion of the
Universe \cite{Lemaitre:1927zz,Hubble:1929ig}. However, while the
precise measurement of the Cosmic Microwave Background Radiation
(CMBR) supplemented by the standard ($\Lambda$CDM) cosmological model
suggests that $H_0 = 67.66 \pm 0.42$
$\mathrm{km} \ \mathrm{s}^{-1}\mathrm{Mpc}^{-1}$ \citep{WMAP:2012nax,Planck:2018vyg},
a model-independent determination using distance ladders
alongwith Type Ia supernovae (SNe) data and the SH0ES project yields $H_0
= 73.04 \pm 1.04$
$\mathrm{km} \ \mathrm{s}^{-1}\mathrm{Mpc}^{-1}$ \cite{Pan-STARRS1:2017jku,Riess:2021jrx}. This
discrepancy (nearly $5 \sigma$) between the outcomes of two profoundly
different measurements (early time vs. local) is termed as the Hubble
tension.

Unless the discrepancy is just an artefact of unresolved systematics,
an explanation would call for some unknown physics
effects~\cite{2020PhRvD.102b3518V}. However, given the internal
accuracy of the data, maintaining the consistency of the redshift ($z$) dependent observable $H(z)$ at both
ends of the distance scale imposes severe constraints on the nature of
the new physics models. This has led to the use of further
model-independent methods for determining $H_0$, an example being the
age of the earliest stellar populations in our
galaxy~\cite{2015ApJ...808L..35T,2019JCAP...03..043J,2020JCAP...12..002V,2021PhRvD.103j3533B,2021MNRAS.505.2764B}. However,
the majority of the age measurements rely on objects at higher
redshifts, which can constrain other cosmological parameters\cite{*1995Natur.376..399B,*1995GReGr..27.1137K,*1996Natur.381..581D,*1999ApJ...521L..87A,*2000MNRAS.317..893L,*2002ApJ...573...37J,*2003ApJ...593..622J,*2004PhRvD..70l3501C,*2005MNRAS.362.1295F,*2006PhLB..633..436J,*2014A&A...561A..44B,*2015AJ....150...35W,*2017JCAP...03..028R,*2020MNRAS.496..888N,*2021ApJ...908...84V}, 
but none serve to resolve the Hubble
tension \cite{DiValentino:2021izs,2022arXiv221104492K,2023Univ....9...94H}.

There are two primary methods for constructing a consistent model of
the universe that would account for both the dynamics of the local
observations, namely the current accelerated expansion of the universe
as well as early universe observations. Einstein's minimal
gravitational action can be modified, and extended gravity theories
constructed \cite{Nojiri:2006ri,DeFelice:2010aj,Capozziello:2011et,
Cai:2015emx}. The alternative is to replace the cosmological constant
by a matter component exerting negative pressure and known as dark
energy
(DE)~\cite{Carroll:2000fy,Peebles:2002gy,Copeland:2006wr,Li:2012dt}. Despite
numerous efforts, there is still no preferred dark energy model that
can precisely describe the universe's dynamical processes, including
the Hubble tension.

Four primary components define the $\Lambda$CDM model:
radiation, dust, scalar curvature, and the cosmological constant. Each
component is barotropic, {\em i.e.} the pressure is proportional to
energy density: $P = w \rho c^2$, where
$w$ is known as the equation of state parameter for the fluid.  For
radiation, $w_r = 1/3,$ whereas $w_m = 0, w_k= -1/3$ and $w_\Lambda =
-1,$ respectively, are the values for dust matter, scalar curvature,
and the cosmological constant. While fits of the $\Lambda$CDM ansatz
with the early \citep{WMAP:2012nax,Planck:2018vyg} as well as
late-time observations \cite{Pan-STARRS1:2017jku,Riess:2021jrx} are
consistent with a vanishing curvature contribution, we do not, {\em a
priori} demand this. Furthermore, to alleviate the tension, we admit
the existence, in the system, of a non-negligible amount of another
barotropic fluid with the equation of state parameter $w_n$. The
Universe was already dominated by the dust matter ($w_m = 0$) during
the recombination era, whereas the current Universe is assumed to be
driven by the cosmological constant $(w_\Lambda = -1)$. Since the
tension arises from the overlap of these two eras, it stands to
reason that any component capable of effecting a significant change in
the dynamics should lie between dust and cosmological constant in its
behavior.  In other words, $-1 < w_n < 0$ is preferred. As the energy
density of the barotropic fluid $\rho \propto a^{-3(1 + w)},$ where
$a(t)$ is the scale factor, it follows that the energy
density of the newly assumed fluid satisfies $\rho_n \propto (1 + z)^n,$
where $0<n<3$.
        
This paper is organized as follows: in 
section \ref{sec_data_method}, we describe the dataset (Pantheon) and
methodology (Markov Chain Monte Carlo) used in this work. The obtained
results, we present in Sec. \ref{sec_results}. Finally,
Sec. \ref{sec_conclu} discusses the main conclusions and summarizes.

%%%%%%%%%%%%%%%%%%%%%%%%%%%%%%%%%%%%%%%%%%
%%% Section 2: Dataset and Methodology %%%
%%%%%%%%%%%%%%%%%%%%%%%%%%%%%%%%%%%%%%%%%%
\section{Dataset and Methodology}\label{sec_data_method}
\subsection{Type Ia Supernovae Dataset}
Type Ia supernovae (SNe) being standard candles,
the luminosity distance measurements for these have been
crucial in establishing the current cosmic acceleration. The
observable of interest is the distance modulus $\mu_{\rm SNe}$ defined as
the difference between the apparent and absolute magnitude
of the individual supernova. This can be related to the observed rest
frame $B$-band peak magnitude $m_B(z)$, the time stretching of the
light curve ($X_1)$ and the SNe color at maximum brightness ($C$)
through~\cite{Pan-STARRS1:2017jku} 
\begin{equation}\label{eq_mu_1}
    \mu_{\rm SNe}(z)=m_B(z)+\alpha\cdot X_1-\beta\cdot C-M_B
\end{equation}
The absolute $B$-band magnitude or the luminosity is almost
independent of the
redshift~\cite{2019A&A...625A..15T,2022JCAP...01..053K}, and is
normally distributed with a very small spread, {\em viz.},
$M_B=-19.220\pm0.042$ \cite{2020PhRvD.101j3517B}.  The nuisance
parameters $\alpha$ and $\beta$ associated with the variables $X_1$
and $C$ have been marginalized for the Pantheon
dataset~\cite{Pan-STARRS1:2017jku} comprising 1048 Type Ia SNe in the
redshift range $0.01\leq z\leq2.26$, and eventually calibrated to be
zero.  Thus, the expression for the observed distance modulus
essentially reduces to just $\mu_{\rm SNe}(z)=m_B(z)-M_B$.

\begin{figure}
    \centering
    \includegraphics[width = 0.48\textwidth]{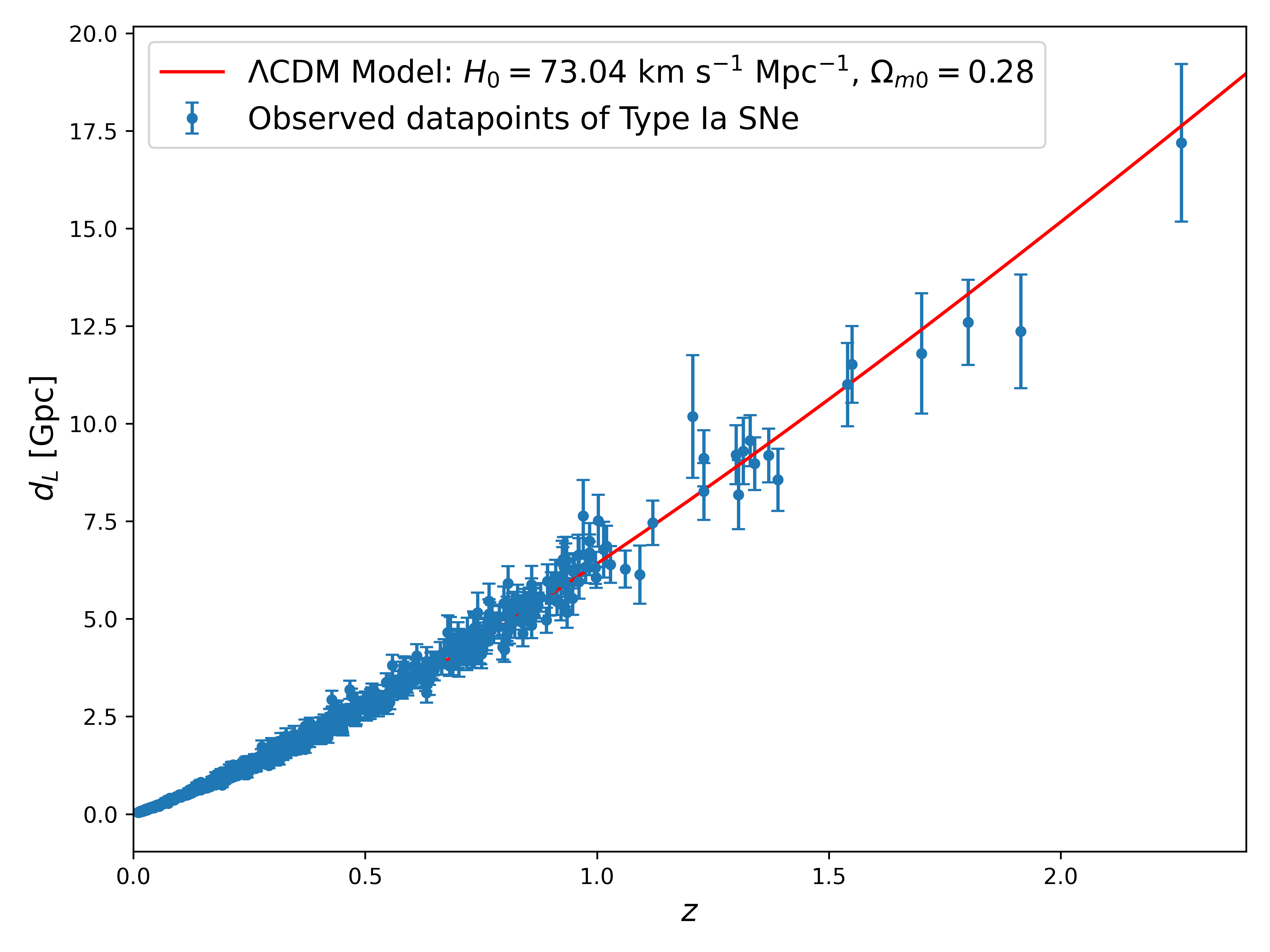}
    \caption{The luminosity distance measurements of Type Ia SNe in
      the redshift range $0.01 \leq z \leq 2.26$. The solid red line
      is the theoretical curve for the flat $\Lambda$CDM model with
      $H_0=73.04$ $\mathrm{km} \ \mathrm{s}^{-1}\mathrm{Mpc}^{-1}$ and
      $\Omega_{m0} = 0.28$.}
    \label{fig_1}
\end{figure}

It is convenient to define a distance
modulus variable $d_{L}(z)$ through  
\begin{equation}\label{eq_mu_2}
   d_{L}^{\mathrm{obs}}\left(z\right)=10^{\left(\mu_{\rm SNe}-25\right) / 5}~~(\text{Mpc}) \ ,
    % \mu_{SNe}(z)=5\log_{10}\left(\dfrac{d_L}{Mpc}\right)+25 \ ,
\end{equation}
with the attendant uncertainty being determined by $\sigma_{m_B}$ alone.
The observations constituting the Pantheon database are depicted in Fig. \ref{fig_1}. 

\subsection{Parameter Estimation}
With the Friedmann equation giving a simple relation between $H(z)$
and the energy density (including that due to a cosmological
constant), it is customary to posit simple equations of state
(barotropic fluids) for the matter components leading to
straightforward expressions for the respective energy densities in
terms of $z$. In the present instant, we invoke an additional
contribution, namely $\Omega_n(z) = \Omega_{n0}(1+z)^n$, where
$\Omega_{n0}$ denotes its value at the current epoch, i.e., at
$z=0$. We have, then,
\begin{eqnarray}\label{eq_5}
H^2(z) &=&  H^2_0 \left[
    \Omega_{r0} (1 + z)^4 + \Omega_{m0}(1+z)^3 +\Omega_{\Lambda}
    \right. \nonumber\\
    &&  \left. \hspace{20pt} +\Omega_{k0}(1+z)^2+\Omega_{n0}(1+z)^n \right],
\end{eqnarray}
where, as usual, $\Omega_{r0},~ \Omega_{m0},$ and $\Omega_{k0}$ are
the current energy density parameters due to radiation, matter, and
curvature respectively. And while $\Omega_{k0} \approx 0$ within the
$\Lambda$CDM, {\em a priori} it is not clear that it would continue to
be so in the revised context. Furthermore, $\Omega_{r0} +
\Omega_{m0}+\Omega_{k0}+\Omega_{\Lambda}+\Omega_{n0}=1$, allowing us
to eliminate one variable in terms of the others, and this we choose
to be $\Omega_{n0}$. The extant data is insensitive to the exact value of
  $\Omega_{r0}$ (since $\Omega_r$ is
already negligibly small by the recombination epoch and plays little
or no role in parameter estimation),
and, hence, for the subsequent numerical analyses, we
  use just the central value {\em viz.} $\Omega_{r0} = 8.5 \times
  10^{-5}$.

Given the expression above for $H(z)$, one can define the luminosity
distance theoretically as

\begin{equation}\label{eq_6}
  d_{L}^{\text{th}}(z;\mathcal{P})=
  \dfrac{c(1+z)}{H_0\sqrt{\Omega_{k0}}}\,\sinh\left[H_0\sqrt{\Omega_{k0}}\int_0^z\dfrac{dz^\prime}{H(z^\prime)}\right] ,
\end{equation}
where $\mathcal{P} \equiv
\{H_0,~\Omega_{m0},~\Omega_{k0},~\Omega_{\Lambda}\}$, $c$ is the speed
of light.

We may now determine the parameters of the model 
by minimizing 
\begin{equation}\label{eq_chisq}
    \chi^2\left(\mathcal{P}\right) = \sum_{i=1}^{N} \left(\dfrac{d_{L}^{\rm th}\left(z_i ; \mathcal{P}\right) - d_{L}^{\text{obs}}\left(z_i \right)}{\sigma_{{d_{L}}}^{\text{obs}}\left(z_i\right)}\right)^2,
\end{equation}
where the sum extends not only over the entire Pantheon database of
Type Ia SNe, but also over distance measurements at the recombination
era.

%%%%%%%%%%%%%%%%%%%%%%%%%%%%%%%%%%%%%%%%%%
%%% Section 3: Results                 %%%
%%%%%%%%%%%%%%%%%%%%%%%%%%%%%%%%%%%%%%%%%%
\section{Results}\label{sec_results}
Our aim is to fit \eqref{eq_5} for $H(z)$ to the
data in the presence of a nonzero $\Omega_{n0}$ and to investigate
whether a qualitative difference exists as compared to the standard
$\Lambda$CDM ({\em i.e.}, the limit of $\Omega_{n0}=0$). While it
might be tempting to consider $n$ itself also as a
continuous variable, we desist from doing so under the assumption that
its value is decided by the underlying theory, whose structure
we remain agnostic about. Instead, we investigate the consequences for
a few discrete values of $n$. We thus have four free
parameters, namely $H_0$, $\Omega_{m0}$,
$\Omega_{k0}$ and $\Omega_\Lambda$, with $\Omega_{n0}$ determined from
the rest. 

\begin{table}[!h]
\centering
\renewcommand{\arraystretch}{2}
\begin{tabular}[b]{| c | c| c| c| c|}\hline
  Param. & \multicolumn{4}{c|}{Best value [68\% C.L.]}\\
  \cline{2-5} 
  & $n = 0.5$ & $n = 1.0$ & $ n = 1.5$ & $n = 2.5$\\
  \hline \hline
  $x$ & $19427^{+471}_{-502}$ & $19357^{+503}_{-482}$
      & $19291^{+501}_{-492}$ & $19168^{+465}_{-456}$\\ \hline
  $y$ & $1399.7^{+29.6}_{-29.0}$ & $1403.8^{+30.2}_{-30.2}$
      & $1408.0^{+31.8}_{-31.2}$ & $1416.9^{+30.8}_{-30.4}$\\ \hline
  $H_0$ & $72.198^{+0.197}_{-0.191}$ & $72.200^{+0.189}_{-0.187}$ 
        & $72.191^{+0.187}_{-0.186}$ & $72.189^{+0.170}_{-0.173}$\\ \hline
  $\Omega_{m0}$ & $0.268^{+0.006}_{-0.006}$ & $0.269^{+0.006}_{-0.006}$
        & $0.270^{+0.006}_{-0.006}$ & $0.272^{+0.006}_{-0.006}$\\ \hline
  $\Omega_{k0}$ & $-0.024^{+0.061}_{-0.059}$ & $-0.053^{+0.080}_{-0.079}$
        & $-0.126^{+0.129}_{-0.129}$ & $0.132^{+0.064}_{-0.066}$\\ \hline 
  $\Omega_{\Lambda}$ & $0.491^{+0.247}_{-0.261}$& $0.594^{+0.114}_{-0.117}$
         & $0.634^{+0.065}_{-0.065}$ & $0.673^{+0.027}_{-0.026}$\\ \hline
  $\Omega_{n0}$ &$0.259^{+0.302}_{-0.312}$ & $0.189^{+0.187}_{-0.189}$ 
         & $0.221^{+0.184}_{-0.187}$ & $-0.077^{+0.048}_{-0.046}$\\ \hline
  $\chi^2_\nu$ & $0.9842$& $0.9846$ & $0.9849$ & $0.9859$\\ \hline 

\end{tabular}
\caption{ The best fit values of $x$, $y$, $H_0$, $\Omega_{m0}$,
  $\Omega_{k0}$, $\Omega_{\Lambda0}$  (along with the $68\%$ C.L. intervals)
  for different values of the parameter $n$. Also displayed are
  the deduced value of $\Omega_{n0}$ and the reduced $\chi^2$.}
	\label{tab_results}
\end{table}

To this end, we employ the \textbf{\texttt{emcee}}-a Python
implementation of a Markov Chain Monte Carlo \cite{emcee} using the
Pantheon dataset \cite{Pan-STARRS1:2017jku}, augmented by the CMBR
data (at $z \approx 1100$). At such large redshifts, $H(z)$ is
determined primarily by the combination $y \equiv H_0^2\Omega_{m0}$,
and the accuracy of the data leads to $y = 1424.18 \pm 31.37$.  Hence,
instead of $H_0$ and $\Omega_{m0}$, we use the combinations $x \equiv
H_0^2/\Omega_{m0}$ and $y$. For $y$, we naturally use a
Gaussian prior (with obvious parameters). For the rest, we use flat
priors, albeit restricted to a certain interval, {\em viz.}
\begin{equation}
    x \in [8000,22000]\ , \quad 
    \Omega_{k0} \in [-1,1]\ , \quad \Omega_\Lambda \in [0,1] \ .
 \label{priors}
\end{equation} 
The flatness of the priors only reflect our ignorance of the said
parameters. While it is tempting to use a Gaussian prior for $x$ as
well, doing so would be tantamount to unduly favoring the
$\Lambda$CDM model as the individual determination of
$\{H_0, \Omega_{m0}\}$ from the CMBR data that such a course would imply
is implicitly dependent on the said model.

Before we discuss the individual cases, we summarise, in
Table \ref{tab_results}, part of the results of our MCMC analysis.
This immediately allows us to infer several results:

\begin{itemize}
\item For all $n$, the obtained best fit value of $H_0$
(see Table \ref{tab_results}) is consistent, within $1\sigma$, with
that determined by the SH0ES survey. That these values are
virtually independent of $n$, signals a robustness of the
fit. This constitutes a central result of our analysis.

\item That the fitted values of $\Omega_{m0}$ (again, virtually
independent of $n$)  are somewhat lower than that in
the $\Lambda$CDM was to be expected and is but a consequence of the
presence of the exotic matter.

\item With the data from the CMBR observations playing an
important role in the fitting and with the combination $y =
H_0^2 \Omega_{m0}$ largely determining the rate of expansion at that
epoch, it was expected that the fitted value of $y$ would be
close to that in $\Lambda$CDM and have only a tiny dependence on $n$
(see Table \ref{tab_results}). This, in turn, also indicates why the
increase in $H_0$ ({\em vis-\`a-vis} $\Lambda$CDM) needs to be
accompanied by a decrease in $\Omega_{m0}$.

\item The dependence on $n$ is, however, very pronounced for the
best fit values of both $\Omega_\Lambda$ and
$\Omega_{k0}$ (and, thereby, $\Omega_{n0}$). In particular, $\Omega_\Lambda$
deviates strongly from the $\Lambda$CDM value of $0.685$ as $n$ deviates from 3.
This is understandable, for $n \to 3$ signals ordinary pressureless
matter (dust), a situation indistinguishable from the $\Lambda$CDM. As $3-n$
increases, the new component becomes more and more exotic and contributes
to suppress $\Omega_\Lambda$.

\item Moving to a discussion for specific values of $n$, we start with
$n = 0.5$. The best fit point for $\Omega_{k0}$
points only marginally towards an open universe, with a flat universe
being as probable. Similarly, the error bars on both
$\Omega_{\Lambda}$ and $\Omega_{n0}$ are relatively large. This is not
unexpected as $n = 0.5$ inidicates an exotic matter that is almost
cosmological constant-like. Consequently, the split between the two
components is not very clearly demarcated. And while it is tempting to
conclude that both $\Omega_{n0}$ and $\Omega_{k0}$ are consistent with
vanishing values, note that the parameters are correlated and
such a conclusion is not well supported (consider, for example, the departure
of $\Omega_{m0} + \Omega_\Lambda$ from unity).

\begin{figure}[!ht]
    \centering
    \includegraphics[width=0.5\textwidth]
    {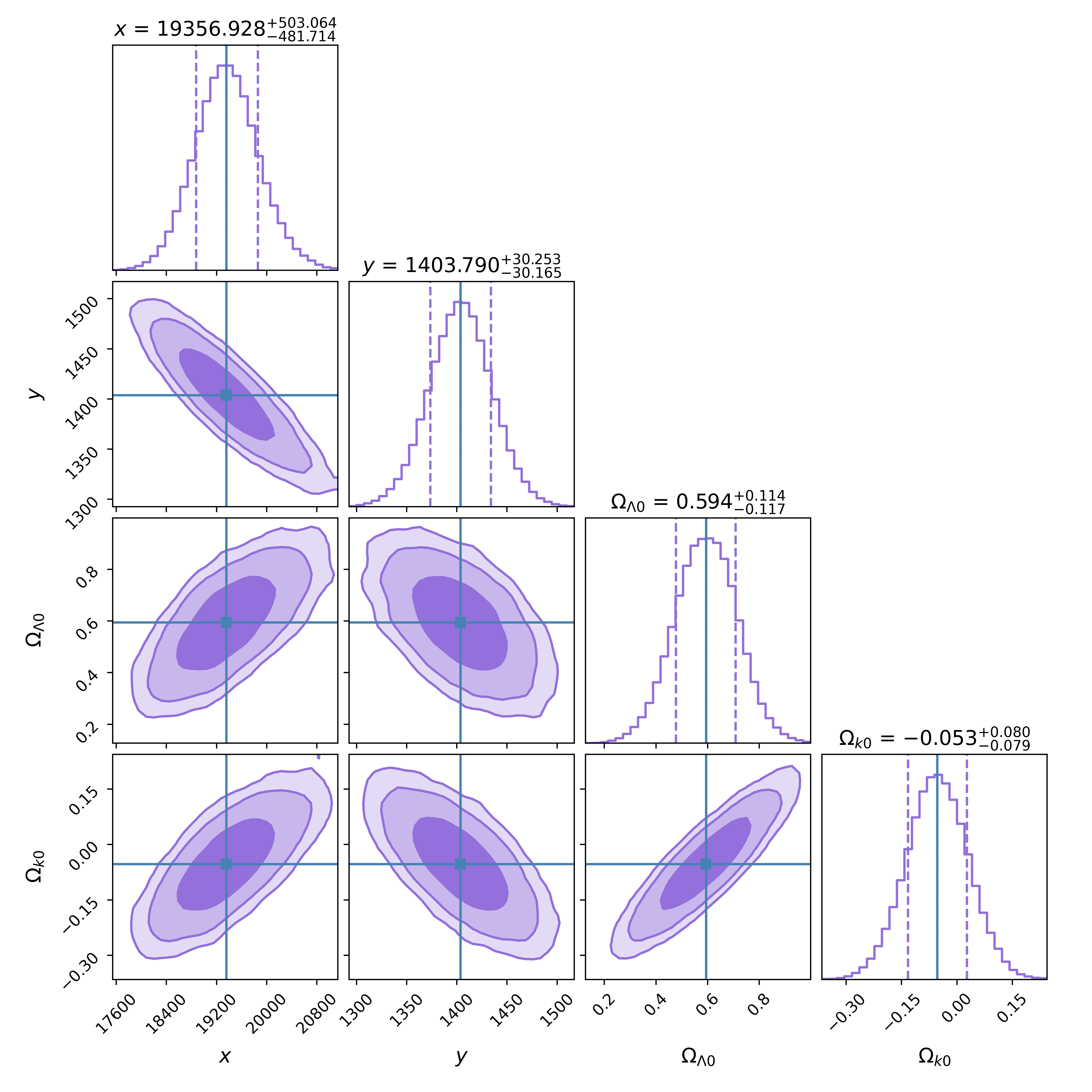}
    \vspace*{-3ex}
    \caption{The 1D and 2D posterior
    distributions of $x$, $y$, $\Omega_{k0}$, and $\Omega_{\Lambda0}$
    for $n = 1$.}  \label{fig:n=1}
\end{figure}

\item For $n = 1$, the posterior distributions for the
parameters are shown in Fig. \ref{fig:n=1}. The preferred value for
$\Omega_{k0}$ has grown more negative (open universe), although a flat
universe is still consistent within $0.7\sigma$. The substantial shift
in $\Omega_\Lambda$ (as compared to the $n = 0.5$ case) is accompanied by a
compensatory shift in $\Omega_{n0}$ (and, to a smaller extent, that in
$\Omega_{k0}$). These features are
not surprising as an $n = 1$ component could be imagined to be midway
between a cosmological constant and a curvature
contribution. Consequently, shifts in the latter would compensate for
any deleterious effects due to $\Omega_{n0}$. Finally, as Fig. \ref{fig:n=1}
shows, the parameters are indeed highly correlated.

\item For $n = 1.5$, the posteriors remain qualitatively
similar, but are sharper. There is, now, a
substantial preference for an open universe, with a flat
universe just about being allowed at $1\sigma$. While $\Omega_{n0}$
has decreased, the preferred value of $\Omega_\Lambda$ has inched
close to the $\Lambda$CDM value. This 
is a consequence of the fact that 
the new component now
behaves quite differently from a cosmological constant and, rather, is
more comparable to the curvature energy. This implies an
anticorrelation between $\Omega_{n0}$ and $\Omega_{k0}$. The remaining
effect is largely compensated by the small shift in $\Omega_\Lambda$.

\begin{figure}[!ht]
    \centering
    \includegraphics[width=0.5\textwidth] {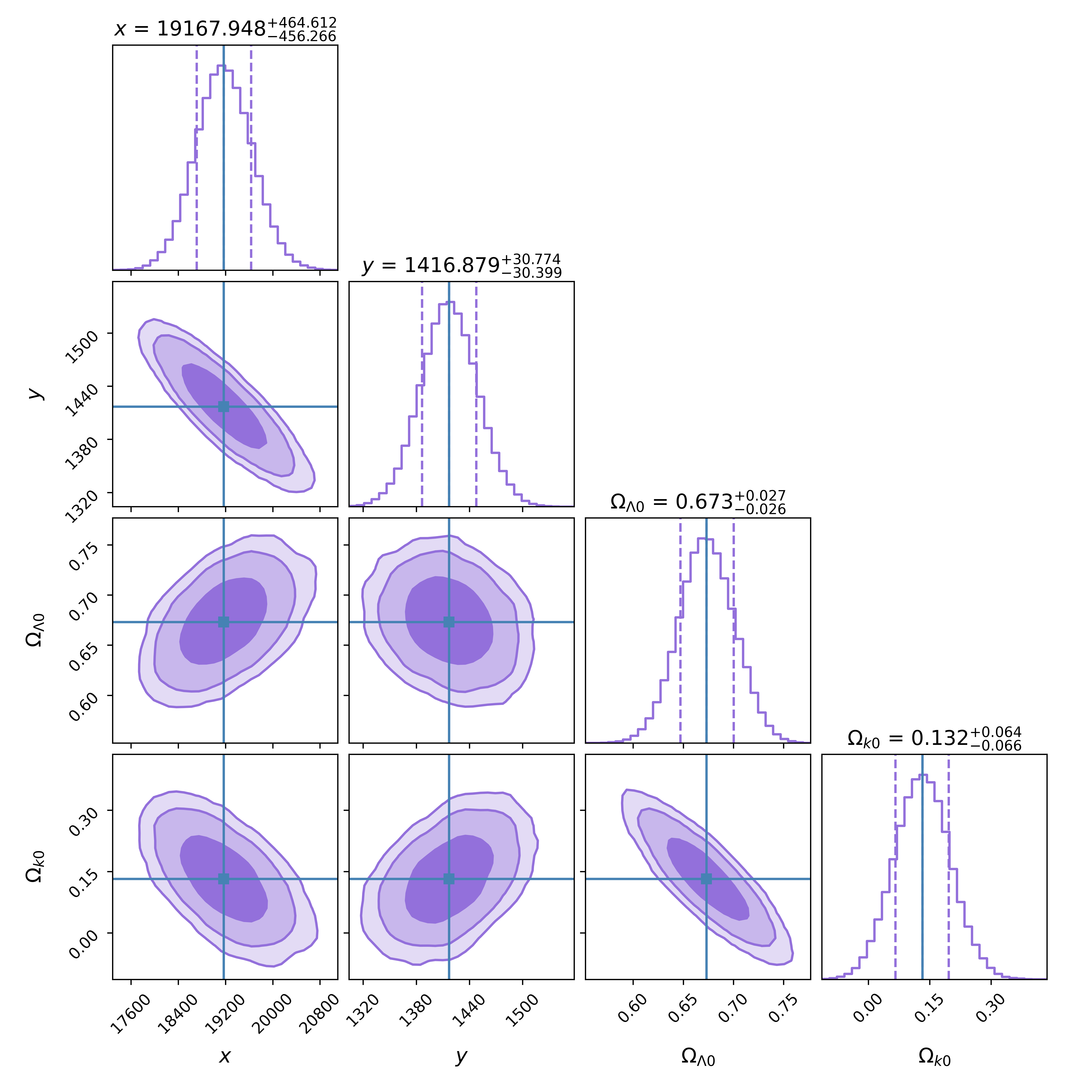}
    \caption{As in Fig. \ref{fig:n=1}, but for $n = 2.5$.}
    \label{fig:n=2.5}
\end{figure}

\item For $n=2.5$, a qualitative difference (as compared to the preceding cases)
appears.  $\Omega_\Lambda$ has moved to a value almost
indistinguishable from that in the $\Lambda$CDM model, but with a relatively
larger uncertainty (see Fig.\ref{fig:n=2.5}). On the other
hand, there is strong preference (at over $2\sigma$) for a closed
universe and $\Omega_{n0}$ has now switched signs! To
appreciate this, note that $\Omega_n$ is now very unlike $\Omega_\Lambda$ and
could, instead, be thought of being halfway between $\Omega_k$ and
$\Omega_m$. Nominally, one would have expected a larger compensatory shift in
$\Omega_{m0}$, but for the fact that the combination $y$ is rather
strongly constrained. With this preventing $\Omega_{m0}$ from moving
very substantially, $\Omega_{n0}$ can only be such that its effect on
$H(z)$ could, to an extent, be compensated for by a non-zero $\Omega_{k0}$.
Naturally, the correlations involving $\Omega_{k0}$ have now switched signs.
\end{itemize}

The excellent value of the reduced $\chi^2$ in each case is a
testimony to the goodness of the fits, with very little to
choose between them (see Fig.\ref{fig:BAO}).

%%%%%%%%%%%%%%%%%%%%%%%%%%%%%%%%%%%%%%%%%%
%%% Section 4: Discussion and Conclusions%
%%%%%%%%%%%%%%%%%%%%%%%%%%%%%%%%%%%%%%%%%%

\section{Discussion and Conclusions}\label{sec_conclu}

\begin{figure}[b]
\hspace*{-4em} 
\includegraphics[width= 0.6\textwidth]{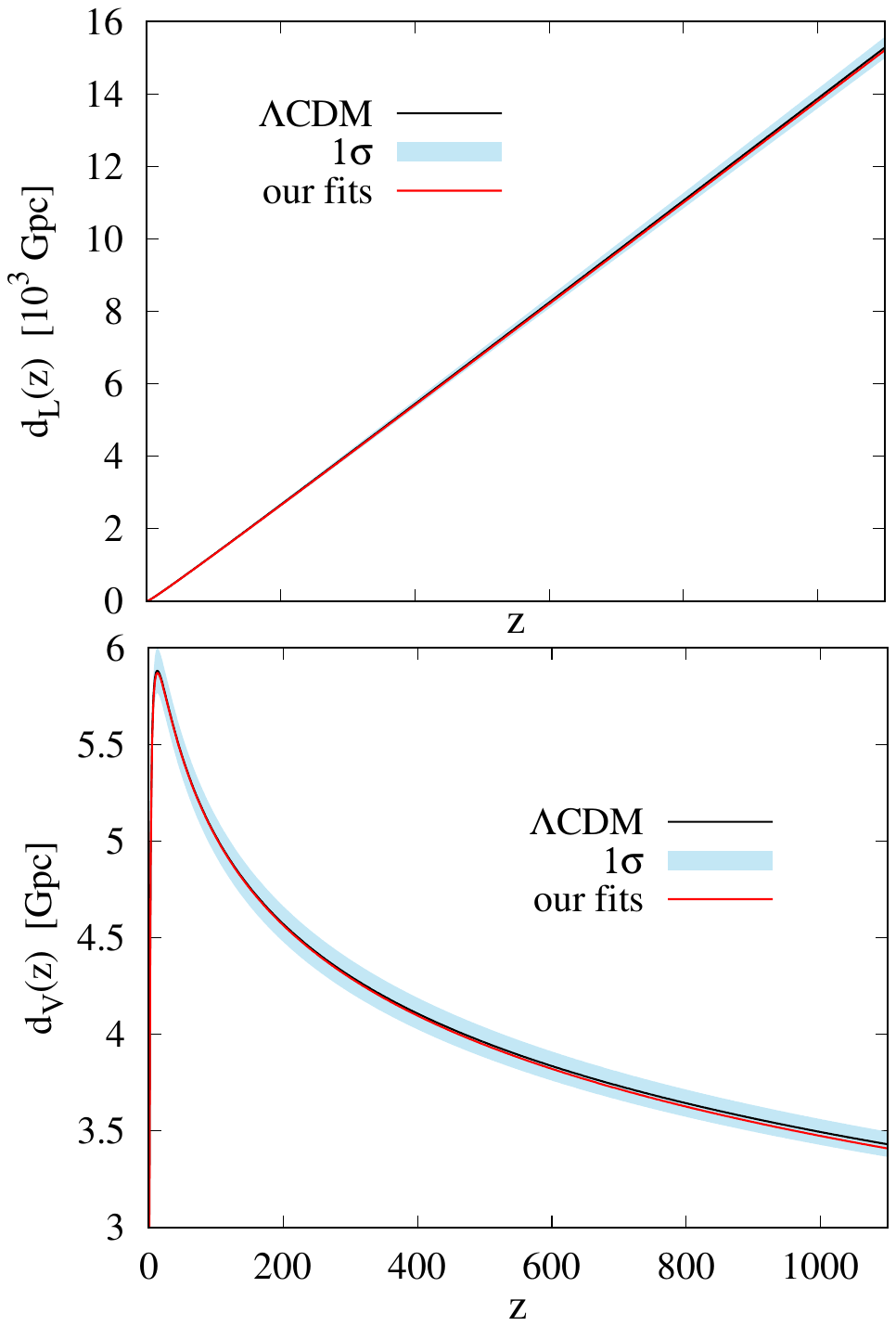}
\vspace*{-16ex}
    \caption{The cosmological distances $d_L(z)$ and $d_V(z)$ as a
    function of the red-shift $z$. Shown are the curves for $n=1$ (on
    the scale of the figures, the others are indistinguishable) as
    well as those for the $\Lambda$CDM model alongwith the
    uncertainties.}  \label{fig:BAO}
\end{figure}
The $\Lambda$CDM model, while
extraordinarily effective at explaining a wide spectrum of
cosmological phenomena, is faced with mounting evidence of a ($\approx
5 \sigma$) discrepancy between the measurements of the Hubble
constant, $H_0$, made locally and that inferred from early universe
observations.  To address this `Hubble tension', we propose a modified
cosmological model with an additional ingredient whose energy density
varies with redshift as $(1 + z)^n$, where $n \in (0,3)$ is a discrete
parameter determined by unknown physics.

The model is confronted with
the Pantheon Compilation, which contains $1048$ data points from Type
Ia Supernovae (SNe), and, simultaneously, required to be consistent
with the $\Lambda$CDM model at redshift $z \approx 1100$.
Regardless of the value of $n$, one of the key findings of the study
is that a value of the present-epoch Hubble constant ($H_0$)
consistent with the published value from the SH0ES survey,
can also be rendered absolutely consistent with the PLANCK analysis.

At this stage, one must acknowledge that the PLANCK analysis is not
the only one sensitive to physics in the early universe. Of increasing
importance in this context are the studies of Baryon Acoustic
Oscillations (BAO). The BAO data analysis incorporates the fitting of
three cosmological distance measures, {\em viz.}, 
\begin{equation}
\begin{array}{rcl}
d_H(z) & \equiv & c / H(z)
\\[1ex]
d_A(z) & \equiv & \left(1+z\right)^{-2} \, d_L(z)
\\[1ex]
d_V(z) & \equiv & \left[z \, (1+z)^2 \, d_H(z) \, d_A^2(z)\right]^{1/3}
\end{array}
\end{equation}
and our projections for these too must agree with the data,
at least as well as the $\Lambda$CDM does.

Noting that $d_A$ and $d_L$ are essentially the same, and that $d_H$
can be obtained from $(d_L, d_V$), in Fig. \ref{fig:BAO}, we
display this pair.
Clearly, for each of the distance measures, our model (for all of the
$n$ values) does as well as the $\Lambda$CDM. In other words, there is
nothing to choose between them as far as the BAO data is
concerned. This is not surprising, for our fitting did take into
account both the low-$z$ and the CMBR data.

%%%%%
Bolstered by this finding, we now return to a recounting of the key
findings.  Scenarios with $n < 2$
show a slight preference for an open universe with a cosmological
constant that is decidedly smaller than that in the $\Lambda$CDM, with the
two departures from the canonical scenario working opposite to
each other at late times. However, it should be remembered that these
scenarios are largely consistent with a vanishingly small $\Omega_{k0}$
and removing this component entirely would not
substantially affect the fit. For $n > 2$, however, a distinct
preference for a closed universe is shown, the effect being most
significant for $2.5\lesssim n \lesssim 2.75$ (with the system
understandably reverting to $\Lambda$CDM as $n \to 3-$).  This
preference for a closed universe is not accompanied
by an increase in $\Omega_\Lambda$ beyond the standard value (although
the values are distinctly larger than those for $n < 2$). Rather, the
effect of the closure is partially compensated for by $\Omega_{n0}
< 0$, i.e., a negative energy density for the exotic matter!  This
invites a deeper analysis of its nature.

Last but not least, the 1D and 2D posterior contour graphs reveal weak
correlations between the parameters. This implies that there may be
subtle interdependencies among the modified model's parameters and
that additional research is required to completely comprehend the
model's behavior and pave the way for insight into the nature of dark
energy and the expansion of the universe.

%%%%%%%%%%%%%%%%%%%%%%%%%%%%%%%%%%%%%%%%%%
%%% Section 5: Acknowledgments    %%%%%%%%
%%%%%%%%%%%%%%%%%%%%%%%%%%%%%%%%%%%%%%%%%%
\section*{Acknowledgments}
We acknowledge the facilities provided by the IUCAA Centre for Astronomy
Research and Development (ICARD), University of Delhi.  DK is supported by an
INSPIRE Fellowship (number IF180293 [SRF]) of the DST, India. DC
acknowledges research grant No. CRG/2018/004889 of the SERB, India. DN is supported by the DST, Government of India through the DST-INSPIRE Faculty fellowship (04/2020/002142).

% \bibliographystyle{apsrev4-1}
% \bibliography{references}


\end{document}